# Quantum plasmonic N00N state in a silver nanowire and its use for quantum sensing

Yang Chen,[1,2] Changhyoup Lee,[3] Liu Lu,[4] Di Liu,[1,2] Yun-Kun Wu,[1,2] Lan-Tian Feng,[1,2] Ming Li,[1,2] Carsten Rockstuhl,[3,5] Guo-Ping Guo,[1,2] Guang-Can Guo,[1,2] Mark Tame,[6,7,*] and Xi-Feng Ren[1,2,*]

[1] *Key Laboratory of Quantum Information, University of Science and Technology of China, CAS, Hefei 230026, China*
[2] *Synergetic Innovation Center of Quantum Information and Quantum Physics, University of Science and Technology of China, Hefei, 230026, China*
[3] *Institute of Theoretical Solid State Physics, Karlsruhe Institute of Technology, 76131 Karlsruhe, Germany*
[4] *School of Mechanical Engineering, Jiangsu University, Zhenjiang 212013, China*
[5] *Institute of Nanotechnology, Karlsruhe Institute of Technology, 76021 Karlsruhe, Germany*
[6] *School of Chemistry and Physics, University of KwaZulu-Natal, Durban 4001, South Africa*
[7] *National Institute for Theoretical Physics (NITheP), KwaZulu-Natal, South Africa*
*Corresponding author: markstame@gmail.com, renxf@ustc.edu.cn*

**The control of quantum states of light at the nanoscale has become possible in recent years with the use of plasmonics. Here, many types of nanophotonic devices and applications have been suggested that take advantage of quantum optical effects, despite the inherent presence of loss. A key example is quantum plasmonic sensing, which provides sensitivity beyond the classical limit using entangled N00N states and their generalizations in a compact system operating below the diffraction limit. In this work, we experimentally demonstrate the excitation and propagation of a two-plasmon entangled N00N state ( $N=2$ ) in a silver nanowire, and assess the performance of our system for carrying out quantum sensing. Polarization entangled photon pairs are converted into plasmons in the silver nanowire, which propagate over a distance of 5 $\mu m$ and re-convert back into photons. A full analysis of the plasmonic system finds that the high-quality entanglement is preserved throughout. We measure the characteristic super-resolution phase oscillations of the entangled state via coincidence measurements. We also identify various sources of loss in our setup and show how they can be mitigated, in principle, in order to reach super-sensitivity that goes beyond the classical sensing limit. Our results show that polarization entanglement can be preserved in a plasmonic nanowire and that sensing with a quantum advantage would be possible with only moderate loss present. © 2018 Optical Society of America**

## 1. INTRODUCTION

Plasmonic systems exploiting quantum optical effects have recently opened up a wide range of devices and applications based on the control of light at the nanoscale [1] including the transmission and generation of entanglement [2,3], single-photon sources [4-7], quantum logic gates [8, 9], quantum random number generators [10], and photonic switches and transistors [11-13]. A vital ingredient in many of these examples is the use of plasmonic waveguides, which enable the realization of highly compact quantum optical circuitry [14, 15]. The nanowire geometry in particular offers several favorable properties for realizing plasmonic waveguides, such as a high confinement of the light field for coupling to emitters [16], e.g. quantum dots [17] and nitrogen-vacancy centers [18], the support of modes with orthogonal polarization [19], and the ability to provide novel types of hybrid modes when combined [20]. While nanowires have been considered for many different quantum applications, a prime example is quantum plasmonic sensing [21-27]. Sensing with plasmonic entanglement will be very useful for fragile systems like photosensitive biological samples, where simply increasing the optical power in order to improve sensitivity would cause optical damage to specimens under investigation [28-33]. Here, it has been shown that by using specially tailored entangled quantum states, called N00N states, and their generalizations, together with carefully chosen measurements, one can achieve optical sensing below the diffraction limit with a sensitivity beyond that of the classical shot-noise limit [24]. Despite the theoretical progress made so far on quantum plasmonic sensing and the potential for exploiting quantum effects, there has not yet been any experimental demonstration of the propagation of a N00N state in a plasmonic nanowire nor a rigorous analysis of its use for quantum sensing, which requires the quantitative identification of losses.

In this work, we experimentally demonstrate the propagation of a two-plasmon entangled N00N state ($N=2$) in a silver nanowire and carry out a detailed analysis to assess the performance of the system for quantum sensing. We use a tapered-fiber silver nanowire hybrid structure to convert polarization entangled photon pairs into plasmons, observing their co-propagation in the nanowire over several micrometers and the re-conversion of the entangled state back into photons. We find that high-quality entanglement is preserved throughout the photon-plasmon-photon transfer process. We then investigate the potential of the N00N state to be used for quantum plasmonic sensing. We measure the characteristic super-resolution phase oscillations expected for the entangled state using coincidence measurements. Due to the presence of loss at specific stages in our setup it is not possible for us to fully demonstrate quantum plasmonic sensing using the N00N state. However, we identify the sources of loss and outline how our setup could be improved in the future in order to reach super-sensitivity to go beyond the classical shot-noise limit. Compared to previous work on the transmission of only one of an entangled pair of photons through a nanowire [19], we demonstrate that both photons of a polarization entangled state can be simultaneously propagated in the nanowire, which requires a protection of the fragile phase relation or quantum coherence between correlated two-photon states. Our work highlights future possibilities for using plasmonic nanowires in the quantum regime via the polarization degree of freedom. It also opens up new directions for experimental work on quantum plasmonic sensing with multi-photon quantum states and specialized measurements.

## 2. EXPERIMENTAL SETUP

The experimental setup used to excite a plasmonic N00N state in a silver nanowire is shown in Figure 1. Here, pairs of photons are generated using type-I spontaneous parametric down conversion [19, 34] (SPDC). We pump a type-I BBO crystal with a 404 nm continuous-wave laser, and use 3 nm narrow bandwidth interference filters (centered at a wavelength of 808 nm) on the photons in each output arm to increase the spectral purity, and suppress background noise from the pump and uncorrelated photon pairs, consequently stretching the coherence length of the photons up to about 200 $\mu m$ (see Figure 1a). The polarizations of the generated photon pairs are adjusted to be aligned along orthogonal directions (horizontal, $|1_H\rangle$, and vertical, $|1_V\rangle$) by half-wave plates, HWP1 and HWP2, respectively. Here, $|n_k\rangle$ denotes a state in a spatial mode consisting of $n$ photons with polarization $k$. The photons from a pair are combined via a polarizing beamsplitter (PBS) and put in the same spatial mode. This operation is followed by HWP3 which is set at $\pi/8$ in order to transform the polarizations such that $|1_H\rangle \to \frac{1}{\sqrt{2}}(|1_H\rangle + |1_V\rangle)$ and $|1_V\rangle \to \frac{1}{\sqrt{2}}(|1_H\rangle - |1_V\rangle)$. Indistinguishability of photon pairs causes a destructive interference in the probability amplitudes that yield the state $|1_H,1_V\rangle$ ($\equiv |1_H\rangle \otimes |1_V\rangle$ with both photons in the same spatial mode), resulting in the polarization-entangled two-photon N00N state written as $|N00N\rangle = \frac{1}{\sqrt{2}}(|2_H,0_V\rangle - |0_H,2_V\rangle)$ being present in the single spatial mode directly after HWP3. This state is then coupled into a single-mode fiber (SMF), with HWP4 used to fine tune the polarization.

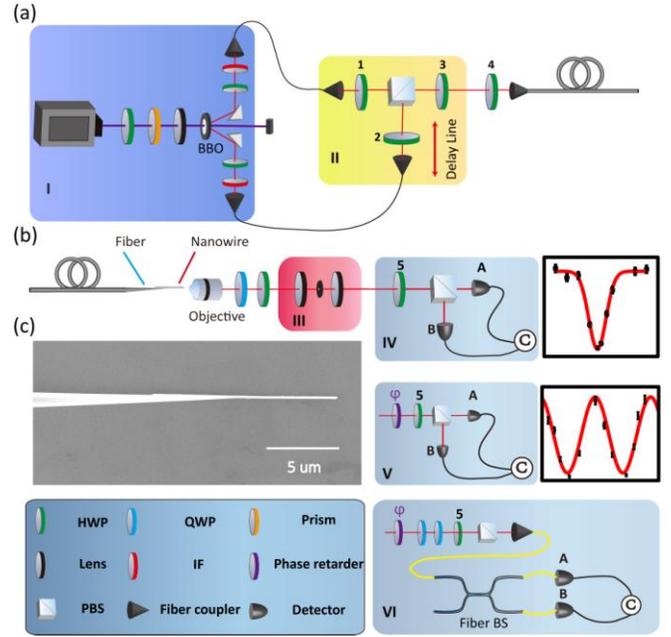

**Fig. 1** Experimental setup. (a) A type-I BBO crystal is pumped by a 404 nm continuous-wave laser to produce two-photon pairs (in region I). The two half-wave plates (HWP1 and HWP2) are used to set each photon of a given pair in an orthogonal polarization. The photons are then put into the same spatial mode using a polarizing beamsplitter. The interference between photons occurs at HWP3 by setting its angle to $\pi/8$. This generates the entangled two-photon state $\frac{1}{\sqrt{2}}(|2_H,0_V\rangle - |0_H,2_V\rangle)$ in the same spatial mode. A separable two-photon state $|1_H,1_V\rangle$ can also be prepared by setting the angle of HWP3 to 0. (b) The input state is transmitted, after HWP4 is used to fine tune the polarizations of the photons, through a tapered-fiber silver nanowire structure, and collected by a 100× objective of 3 mm focal length. We also use an additional HWP and QWP after the objective to fine tune the polarization, and a pinhole confocal system to improve the signal-to-noise ratio (in region III). Subsequently, the two-photon HOM interference and the two photon de-Broglie wavelength measurements are implemented by a coincidence measurement (in regions IV and V, respectively), and quantum state tomography (QST) is performed (in region VI) to characterize the transmitted two-photon entangled state. (c) SEM image of the hybrid structure used to convert photons into plasmons. It consists of a tapered single-mode fiber joined to a silver nanowire with a radius of about 160 nm. The scale bar in the lower-right corner is 5 $\mu m$.

Recently, a tapered-fiber nanowire coupled structure has been shown to efficiently convert photons to plasmons [35]. In our experiment, we employ the same structure to excite a plasmonic N00N state in a single nanowire. The tapered fiber is made by stretching the SMF (in Figure 1a), and a taper tip is formed with the smallest diameter of about 60 nm (see Figure 1b). Then, we use this tapered fiber to lift the silver nanowire with 320 nm diameter. The coupling length between the tapered fiber and the silver nanowire is about 7 $\mu m$, achieving an efficient mode coupling (see Supplement 1). This enables the conversion of the N00N state from photons to plasmons, which then propagate along the silver nanowire with a length of about 5 $\mu m$. A scanning electron microscope (SEM) image for the tapered-fiber nanowire structure is shown in Figure 1c.

As shown in Figure 1b, we use a $100\times$ objective lens ($NA = 0.9$) to collect the light in the plasmonic modes from the end of the nanowire and reconvert it to photonic modes in free space. Afterwards, a 25 $\mu m$ diameter pinhole and two 35 mm focal-length lens are used as a confocal system (region III of Figure 1b) to improve the signal-to-noise-ratio (SNR) by collecting light from the nanowire tip only. The SNR of this system is measured to be 4.8:1 when collecting light just after the confocal system. SMFs are also used at the detection stage to further spatially filter the light and improve the SNR, and the scattered noise can be almost neglected. The noise of the system includes the scattered light from tapered fiber, the silver nanowire, and background noise.

## 3. RESULTS AND DISCUSSION

One requirement for plasmonic devices being used in quantum information applications is to have indistinguishable particles [8,36-38], which can be confirmed by a Hong-Ou-Mandel (HOM) experiment [39]. Indeed, recent experiments have made use of this property to probe the decoherence of path-entangled plasmon N00N states [40] and to evaluate the role of losses [41]. While these works have focused on the number state degree of freedom, here we extend the study of indistinguishability to the polarization degree of freedom. Various aspects of quantum states of light are analyzed using different measurement schemes: the two-photon Hong-Ou-Mandel (HOM) interference detection (shown in region IV of Figure 1b), the two-photon de-Broglie wavelength detection (shown in region V of Figure 1b), and quantum state tomography (QST) (shown in region VI of Figure 1b). These measurement schemes enable us to fully evaluate properties of both the input and output states considered. Further details of these measurements and analysis of the N00N state's interference potential for sensing purposes are described next.

### A. HOM interference
We first evaluate the degree of indistinguishability of the generated two photons by measuring HOM interference. This is achieved by using coincidence detection and a PBS, as shown in region IV of Figure 1b, which is connected to the output of region II in Figure 1a. The relative arrival time of the two photons is varied by using the delay line in region II, and HWP3 and HWP5 are set at 0 and $\pi/8$, respectively. HWP3 has no effect at this stage and the output state from region II is $|1_H,1_V\rangle$, while HWP5 and the PBS in region IV play the role of a 50/50 beamsplitter in a standard HOM setup. The HOM interference is measured yielding a visibility of $0.950 \pm 0.012$ obtained from a fitted curve (see Supplement 1). All the experimental data are raw data without background and accidental subtraction. This clearly shows that the generated two photons that are to be injected into the tapered-fiber silver nanowire structure are highly indistinguishable.
To see if the initial indistinguishability is preserved when the two photons are transferred through the tapered-fiber silver nanowire hybrid structure, we perform the coincidence measurement in region

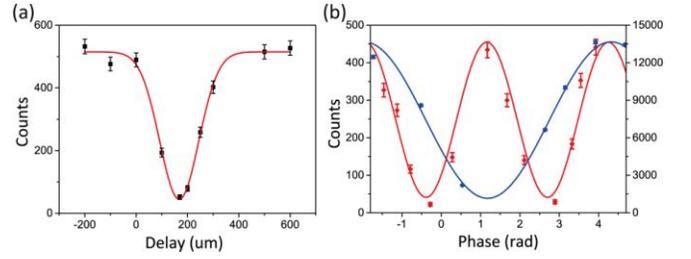

**Fig. 2** (a) Fitted HOM interference curve for the two-photon state transmitted through a silver nanowire. Black squares represent the two photon coincidence counts in 10s. Error bars added to the actual data are determined using a Poisson error in the detection due to the probabilistic nature of the SPDC process being the dominant source of fluctuation in counts [19, 34]. (b) Two photon coincidence counts and single photon counts of the plasmonic N00N state transmitted through a silver nanowire plasmonic system. Red squares represent the two photon coincidence counts in 5s, while blue dots represent the single photon counts in 1s. Here, a phase shift is applied to the single photon case to align the two sets of data. Error bars correspond to Poisson noise due to the SPDC process [19, 34].

IV for the collected photons from the objective lens. In Figure 2a, we present the measured HOM coincidence curve with a visibility of $0.913 \pm 0.028$. Therefore, both the generated two photons and the transmitted two photons through the hybrid structure are highly indistinguishable.

### B. Two-photon de-Broglie wavelength measurements
We also measure the two-photon De-Broglie wavelength [42] using the setup shown in region V of Figure 1b, where the phase retarder accumulates the relative phase $\phi$ between the orthogonal polarization modes of a given state and HWP5 is set at $\pi/8$ in order to transfer this phase change into an amplitude change. Here, the delay line in region II is fixed to 0. The whole structure from region II to region V is equivalent to a Mach-Zehnder interferometer (MZI) in the orthogonal polarization basis.

In the single-photon input case, where the initial photon $|1_V\rangle$ is blocked, the input state $|1_H\rangle$ evolves to $\frac{1}{\sqrt{2}}(|1_H\rangle + e^{i\phi}|1_V\rangle)$ inside the MZI (with HWP3 set to $\pi/8$), ignoring an undetectable overall phase. We measure only one output port of the second PBS (here the port for the horizontally polarized state) and the probability of a single-photon detection is given as

$$P_{|1_H,0_V\rangle}(\phi) = \frac{f_1(1+V_1\cos\phi)}{2}, \quad (1)$$

where the overall factor $f_1$ and the single-photon visibility $V_1$ depend on losses (or efficiencies). On the other hand, when two photons are injected into the MZI in orthogonal polarizations, generating the two-photon N00N state $\frac{1}{\sqrt{2}}(|2_H,0_V\rangle - |0_H,2_V\rangle)$ inside the MZI, the probability of a coincidence count is given as

$$P_{|1_H,1_V\rangle}(\phi) = \frac{f_2(1+V_2\cos 2\phi)}{2}, \quad (2)$$

where the overall factor $f_2$ and the two-photon visibility $V_2$ also depend on losses (or efficiencies) and the phase $\pi$ from the second term of the N00N state has been absorbed into the accumulated phase $2\phi$. Note that the overall factors $f_1$ and $f_2$ are the total proportion of photons that lead to a single-photon detection and two-photon coincidence count, respectively.

We first measure the single-photon and two-photon interference for the respective photonic input states in the absence of the tapered-fiber nanowire hybrid structure. The single-photon detection and two-photon coincidence counts are observed while altering the applied voltage of the phase retarder, showing an oscillation with $\phi$ (see Supplement 1). From the fitted curves, the rapid oscillation in the two-photon case, which is twice as fast as the single-photon case, is demonstrated. The measured visibilities are $0.953 \pm 0.003$ and $0.928 \pm 0.003$ for the single-photon and two-photon case, respectively. We perform the same experiment for the states transmitted through the tapered-fiber structure, but without the nanowire. The measured visibilities of the single-photon and the two-photon interferences are $0.991 \pm 0.004$ and $0.898 \pm 0.015$, respectively. The oscillation frequency of the two-photon case is again twice that of the single-photon case (see Supplement 1).

Finally, we carry out the measurement for the states transmitted through the tapered-fiber nanowire hybrid structure. Our novel scheme to convert a quantum state of light between a tapered fiber and a silver nanowire involves a polarization-dependent coupling efficiency due to the azimuthal asymmetry of our hybrid structure. For the two-photon N00N state $\frac{1}{\sqrt{2}}(|2_H, 0_V\rangle - |0_H, 2_V\rangle)$ being injected, the N00N state transmitted through the silver nanowire can be written in normalized form as $\alpha_2 |2_H, 0_V\rangle - \beta_2 e^{i\phi_0} |0_H, 2_V\rangle$, where $\alpha_2^2 + \beta_2^2 = 1$, and the relative phase $\phi_0$ is considered as a result of the birefringence effect that might occur in the non-symmetric geometry of the coupling region, which can be absorbed into the accumulated phase $\phi$. The ratio of normalized two-photon coupling efficiencies between $H$ and $V$ polarizations for the input state $\frac{1}{\sqrt{2}}(|2_H, 0_V\rangle - |0_H, 2_V\rangle)$ is measured to be $\alpha_2 : \beta_2 = 1.65 : 1$. While this imbalance does not affect $f_2$ in Eq. (2) (once it is renormalized), the expected two-photon visibility is affected and given by $V_2 = 2\alpha_2\beta_2 = 0.887$. The corresponding expected single-photon visibility is $V_1 = 0.969$. In Figure 2b, the measured oscillation with $\phi$ is shown for both the single-photon input and two-photon N00N state input, showing visibilities of $0.737 \pm 0.007$ and $0.880 \pm 0.013$, respectively. We can clearly see the expected change in the oscillation period due to the N00N state. Both values are below their expected theory ones, however, which may be due to the non-ideal states generated in the experiment, background counts at the detectors and to a lesser extent, unwanted polarization mixing at the waveplates and PBS.

## C. Quantum state tomography

To evaluate the presence of entanglement in both the input and output N00N state of the hybrid nanostructure, quantum state tomography (QST) is performed (as shown in region VI of Figure 1b. The reconstruction has been carried out according to the quantum state tomography procedure outlined in Dieleman et al. [3]. Here, the coincidence counts from nine measurement settings of the wave plate angles enable a full characterization of the state in the two-photon subspace. We quantify the closeness of the experimental state to the ideal state using the fidelity, defined as $F = \langle \psi | \rho | \psi \rangle$, where $|\psi\rangle = \frac{1}{\sqrt{2}}(|2_H, 0_V\rangle - |0_H, 2_V\rangle)$ is the ideal state and $\rho$ is the experimental state of the renormalized two-photon subspace, obtained from the tomography.

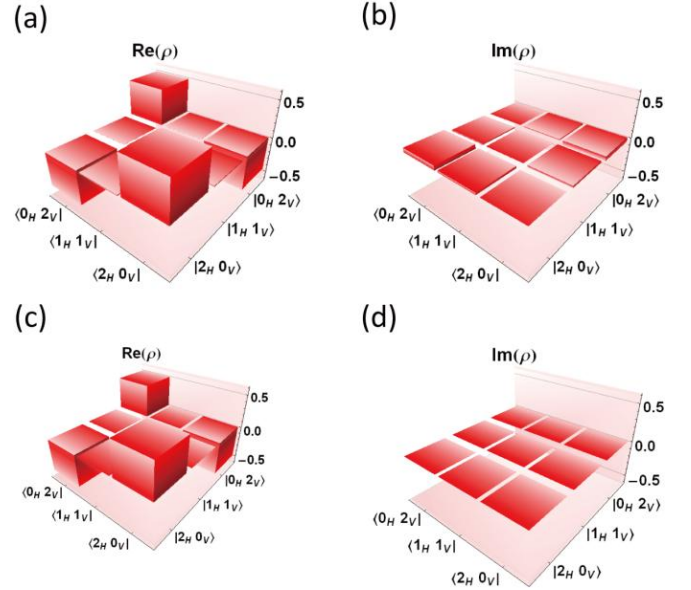

**Fig. 3** Quantum state tomography of the N00N state input ($N = 2$) in the two-photon subspace of the Hilbert space. (Upper row) Experimentally reconstructed state: (a) the real part and (b) the imaginary part. (Lower row) Ideal state: (c) the real part and (d) the imaginary part.

The state initially generated in regions I and II consists of a weighted superposition of different photon number polarization states $|n_H, m_V\rangle$ in the same spatial mode. During the conversions between free space, the fiber, and the nanowire, imperfect conversion efficiencies and transmission losses eventually redistribute the photon number statistics of the state. Since our work focuses on the generation of a two-photon N00N state in the plasmonic nanowire, the state tomography is conducted for the two-photon subspace of the Hilbert space. The photon number states $|n_H, m_V\rangle$ for $n_H + m_V < 2$ do not contribute to the two-photon subspace and higher photon number states have negligible effect on the two-photon subspace at the pump powers used [3]. A rigorous examination is needed to determine whether quantum coherence in the two-photon subspace is preserved during the multiple conversion processes, i.e. whether the output state is still the same as the ideal two-photon N00N state.

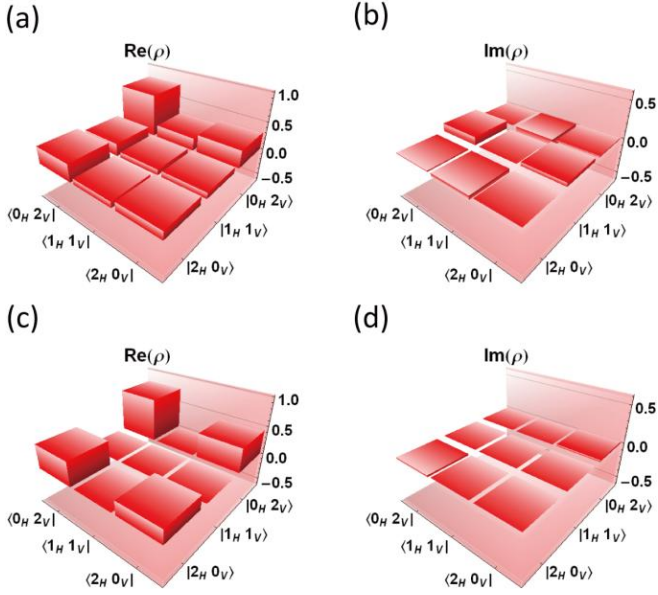

**Fig. 4** Quantum state tomography of the unbalanced N00N state output ($N=2$) transmitted through a silver nanowire in the two-photon subspace of the Hilbert space. (Upper row) Experimentally reconstructed state: (a) the real part and (b) the imaginary part. (Lower row) Ideal state: (c) the real part and (d) the imaginary part.

In Figure 3 we show the reconstructed quantum state of the two-photon N00N state input in the two-photon subspace of the Hilbert space. A fidelity of $F_{in} = 0.957 \pm 0.018$ is obtained when comparing the experimental state with the ideal two-photon N00N state, indicating a high quality of the prepared two-photon N00N state to be injected into the nanowire (ideally $F=1$). In Figure 4 we show the reconstructed quantum state output from the nanowire in the two-photon subspace of the Hilbert space. We obtain a fidelity of $F_{out} = 0.879 \pm 0.040$, showing that the experimental state is close to the expected one. The expected state in this case is defined by choosing the optimal phase and amplitudes of the ideal output state to maximize the fidelity. The chosen optimal phase $\phi = 3.20$ rad and the ratio of amplitudes $\alpha_2 / \beta_2 = 1.78$ are consistent with the measurements from the previous section. The expected state is shown in Figure 4. The lower bound on the entanglement of the experimental output state, as quantified by the concurrence [43], is $C = 0.639 \pm 0.058$, showing that a high degree of entanglement is present in the state (ideally $C=1$). Here, errors have been calculated using a Monte Carlo simulation with Poissonian noise on the measured coincidence counts [3, 19, 34].

### D. Super-resolution and super-sensitivity

The silver nanowire is now investigated for its use in quantum sensing for phase estimation. We consider the sensing device as the central interferometric part of the setup, with an input that begins at the PBS in region II in Figure 1a, and ends with its output after the PBS in region V in Figure 1b. We then study the operation of the device between these input and output points. This allows us to obtain a performance of the quantum plasmonic sensing device that is independent of the photon source and detectors [44]. Such an approach is acceptable since we do not consider the whole structure including the source and detector parts as a sensing device, but rather consider the usefulness of the tapered-fiber silver nanowire hybrid structure for quantum sensing. For this purpose, we exclude the detection efficiency $\eta_D$ from Eqs. (1) and (2), (see Supplement 1 for details).

The device we consider is equivalent to a MZI, where orthogonal polarization modes play the role of the usual orthogonal path modes, one as the reference ($H$) and the other as the probe ($V$). Instead of inducing the variation of the relative phase $\phi = \phi_V - \phi_H$ between the orthogonal polarization modes directly in the nanowire, which requires further modification to the setup, we vary the phase using a liquid crystal phase retarder outside the nanowire (as depicted in Figure 1b and explained in Section 3B), so as to gain an understanding of the performance of the device in a more controlled manner. This is completely equivalent to the case that the relative phase $\phi$ is accumulated directly in the nanowire, upon the fact that the phase shift operations commute with any loss operations that may be present [45]. A polarization-dependent phase shift, yielding the relative phase, could be obtained in the nanowire using a ligand coating [46, 47] or active coating [48] along one of the axes that interacts with a substance whose concentration is being sensed. Such a coating can be achieved via a full coating [48-51] and then nanoshaving [52, 53], or directly via nanografting [53, 54].

One aspect that has been often investigated for quantum sensing is super-resolution [55]. Figure 2b clearly shows the super-resolution of the interference fringes for $N=2$, where the interference oscillation occurs over a phase $N$ times smaller than one cycle of classical light (the $N=1$ case and the classical case have an equivalent resolution). It is known that super-resolution can also be achieved by an engineered classical source of light [55] and so an additional aspect has to be investigated for sensors to demonstrate a quantum enhancement in sensitivity. This aspect is super-sensitivity, i.e. whether a sensor can provide phase-sensitivity beyond the classical shot-noise limit. This limit in the classical case is the standard interferometric limit (SIL) and the minimum phase estimation error (phase sensitivity) is given by

$$\Delta \phi_{SIL} = \frac{1}{\sqrt{\eta_{overall}} N}, \quad (3)$$

where $\sqrt{\eta_{overall}} = 2\sqrt{\eta_H \eta_V} / (\sqrt{\eta_H} + \sqrt{\eta_V})$, and $\eta_H$ and $\eta_V$ denote the overall transmittivities of the $H$ and $V$ polarized modes through the MZI device [56] (see Supplement 1 for the experimental identification of losses in our system). For the N00N state input evolved inside the MZI, followed by an $N$-fold coincidence measurement, the probability of a coincidence count is generally written as

$$P_{coin}^{(N)} = \frac{f_N (1 + V_N \cos N\phi)}{2}, \quad (4)$$

where $f_N$ is the proportion of the input state that leads to an $N$-fold coincidence detection event and $V_N$ is the $N$-photon visibility. The phase sensitivity of a measurement of coincidence detection events is lower bounded by the Cramer-Rao bound, written as [57]

$$\Delta \phi = \frac{2\sqrt{\sigma^2}}{f_N V_N N |\sin(N\phi)|}, \quad (5)$$

where the Fisher information has been calculated for the probability distribution of the $N$-fold coincidence detection, $P_{coin}^{(N)}$, and $\sigma^2$ represents the variance of the outcomes, defined as $\sigma^2 = P_{coin}^{(N)} (1 - P_{coin}^{(N)})$. One can then construct an inequality, $\Delta \phi < \Delta \phi_{SIL}$, that must be satisfied for the sensor to be regarded as super-sensitive [55, 58]. Taking into account the worst case of $\sigma^2 = 1/4$ and the point of minimum phase sensitivity, i.e. when $|\sin(N\phi)| = 1$, the inequality becomes

$$1 < \frac{f_N^2}{\eta_{overall}} V_N^2 N, \quad (6)$$

In other words, the measured visibility has to be greater than the threshold visibility defined as $V_{th} = \sqrt{\eta_{overall}/f_N^2 N}$. In our case we have $N = 2$ and in the absence of loss, i.e. $\eta_{overall} = 1$ and $f_2 = 1$, we must obtain a visibility $V_2 > V_{th}$ where the threshold visibility is $V_{th} = 0.707$. Our experimentally measured visibility $V_2$ of $0.880 \pm 0.013$ is clearly above the threshold value, showing super-sensitivity in principle. However, once loss has been included the threshold visibility increases, and when $\eta_H = \eta_V = \eta$ for simplicity, for $\eta$ smaller than 0.794 it is no longer possible to satisfy the inequality. Furthermore, when $\eta_H \neq \eta_V$, the inequality $f_2^2/\eta_{overall} < 0.5$ corresponds to a regime in which it is not possible to achieve super-sensitivity in any way. In our experiment, we identified $f_2^2/\eta_{overall} = 1.23 \times 10^{-2}$ for our plasmonic setup (see Supplement 1 for details), and therefore even if $V_2 = 1$ our setup would not provide super-sensitivity.

More interestingly, the inequality of Eq. (6) can be simplified to $1 < \eta_H \eta_V \left(\sqrt{\eta_H} + \sqrt{\eta_V}\right)^2/2$, for which the two-photon visibility has been substituted in as $V_2 = 2\eta_H\eta_V/\left(\eta_H^2 + \eta_V^2\right)$ (see Supplement 1).

This enables us to find a lower bound for $\eta_H$ and $\eta_V$, above which the 2-fold coincidence measurement for the two-photon state input can lead to super-sensitivity. The inequality can be reduced to approximately $1.6 < (\eta_H + \eta_V)$, which indicates that it would be possible to satisfy this inequality with a few improvements to the setup. For instance, we could improve the coupling strength from the single-mode fiber to the nanowire and reduce the nanowire propagation loss (using a shorter length nanowire while keeping a detectable phase change to be accumulated) to reach a total coupling ratio of 0.8. Furthermore, with improvements to the collection optics and the confocal microscope, values close to $\eta_{obj} = \eta_c \simeq 1$ could also be achieved. Keeping the current coupling strength of 0.883 from free-space to single-mode fiber leads to the setup reaching the regime of super-sensitivity. Obtaining $\eta_{obj} = \eta_c \simeq 1$ may seem demanding, however, improvements to the coupling strength of light from free space to the single-mode fiber would relax the values needed for these efficiencies. Thus, with reasonable improvements to the setup it would be possible in principle to reach super-sensitivity and go beyond the classical shot-noise limit, demonstrating a quantum advantage in sensing even in the presence of loss. This is obviously a fruitful direction for future work, although not without its experimental challenges.

More realistically, other strategies may improve the situation further, such as the use of generalized N00N states and optimized measurements [24]. This would relax the lower bound of the efficiencies $\eta_H$ and $\eta_V$ that need to be reached in the experiment for showing super-sensitivity.

Notably, with the help of a plasmonic system, the footprint of our sensing device is only a few micrometers, which is several orders of magnitude smaller than that in the previous work [44]. This helps if there is only a trace quantity of a substance that needs to be sensed, or the sensing system is part of a larger photonic system that includes a source and detection part, and the footprint needs to be kept to a minimum.

## 3. CONCLUSIONS

We experimentally demonstrated the excitation and propagation of a two-plasmon entangled N00N state in a silver nanowire, and assessed the performance of the system for carrying out quantum sensing. A full analysis of the plasmonic system showed that high-quality entanglement is preserved throughout. We measured the characteristic super-resolution phase oscillations of the entangled state via coincidence measurements and identified the various sources of loss in our setup, showing how they can be improved in order to achieve super-sensitivity. The results show that polarization entanglement can be preserved in a plasmonic nanowire and that sensing with a quantum advantage is possible with moderate loss present.

**Funding.** National Natural Science Foundation of China (NSFC) (Nos.61590932, 11774333, 11204107), the Anhui Initiative in Quantum Information Technologies (No.AHY130300), the Strategic Priority Research Program of the Chinese Academy of Sciences (No.XDB24030600), the National Key R & D Program (No.2016YFA0301700) and the Fundamental Research Funds for the Central Universities.

**Acknowledgment.** M. T. thanks the support of the South African National Research Foundation, the University of KwaZulu-Natal Nanotechnology Platform and the South African National Institute for Theoretical Physics. This work is partially carried out at the USTC Center for Micro and Nanoscale Research and Fabrication.

See Supplement 1 for supporting content.